\documentclass{article}
\def\stareq{\buildrel\ast\over =}
\begin{document}
\title{On the structure of the energy-momentum and the spin currents
  in Dirac's electron theory\thanks{To appear in: {\sl
    On Einstein's Path.  Festschrift for E. Schucking on the occasion
    of his 70th birthday.} A. Harvey, ed. (Springer: Berlin 1997/98)}}
\author{
Friedrich W.\ Hehl\\ Institute for Theoretical Physics, 
     University of Cologne\\ D-50923 K{\"o}ln, Germany \and 
Alfredo Mac\'{\i}as, 
Eckehard W.\ Mielke\\ Departamento de F\'{\i}sica\\ Universidad 
     Aut\'onoma Metropolitana--Iztapalapa\\ P.O. Box 55-534, 
     09340 M\'exico D.F., M\'exico \and 
Yuri N.\ Obukhov\\ Department of Theoretical Physics, Moscow State
     University\\ 117234 Moscow, Russia} 

\date{1997-06-03} 

\maketitle
\begin{abstract}
  We consider a classical Dirac field in flat Minkowski spacetime. We
  perform a Gordon decomposition of its canonical energy-momen\-tum
  and spin currents, respectively. Thereby we find for each of these
  currents a convective and a polarization piece. The polarization
  pieces can be expressed as exterior covariant derivatives of the
  two-forms $\check M_\alpha$ and $M_{\alpha\beta}=-M_{\beta\alpha}$,
  respectively.  In analogy to the magnetic moment in electrodynamics,
  we identify these two-forms as {\it gravitational moments} connected
  with the translation group and the Lorentz group, respectively. We
  point out the relation between the Gordon decomposition of the
  energy-mo\-men\-tum current and its Belinfante-Rosenfeld symmetrization.
  In the non--relativ\-is\-tic limit, the translational gravitational
  moment of the Dirac field is found to be proportional to the spin
  covector of the electron. {\it File schuecking7.tex, 1997-06-03} 
\end{abstract}

\section{Introduction}

Fermi systems do not possess classical analogs. Nevertheless, for such
systems, we can define a limit for $\hbar\rightarrow 0$, even if, in
this limit, we do not have ordinary classical theory. We feel that a
description of such a limit would be interesting in view of the
possibility to develop a better intuition and to build up new models
\cite{casa}. Conventionally Fermi systems can be quantized by the path
integral method. By introducing semi-classical Grassmann variables,
for instance, we get a deeper understanding of this method
\cite{martin}.

In some dual theories with Fermi variables, a detailed knowledge of
the classical limit is also interesting, because it allows a
reinterpretation in terms of superstrings \cite{sch,ram}. This unified
treatment of commuting and anticommuting variables is obviously of
particular interest for supersymmetric theories, because in this case
one is obliged to treat Fermi and Bose fields in a symmetrical way
\cite{sch,ram,nee}.

Knowing the usefulness of understanding the classical limit, we will
investigate, in this article, the structure of the energy-momentum and
the spin current of the {\it classical} Dirac theory. Of course, the
classical Dirac theory can only be made consistent by second
quantization \cite{jost}. However, for low energy phenomena, when
particle creation and annihilation don't play a role, the classical
Dirac theory and its non-relativistic limit convey an appropriate
picture of the underlying physics.

An electron carries a spin of $s_z=\hbar/2$. It is legitimate to
visualize this spin as some sort of intrinsic circular {\it motion}.
This seems to be clear from the relation of spin to the Lorentz or the
three-dimensional rotation group \cite{Wigner}, but it can also be made
explicit by studying the quantum mechanics of a Dirac particle in a
Coulomb potential. Even in an $s_{1/2}$ state, an electric current is
present, i.e., ``a polarized Dirac electron is a {\it rotating}
particle''\cite{Yang}.

If we take this for granted, it is obvious that the electric charge,
which is housed in the electron, induces an Amp\`ere type ring current
which, in turn, according to the Oersted-Amp\`ere law, acts as a
magnetic moment $\mu_{\rm e}$. The specific nature of spin angular
momentum yields a gyromagnetic ratio $\mu_{\rm e}/ (s_z/\hbar)
=-2\mu_{\rm B}$, where $\mu_{\rm B}:=e\hbar/2m_{\rm e}$ is Bohr's
magneton ($e$ is the charge of the electron, $m_{\rm e}$ its mass).
This gyromagnetic ratio turns out to be twice as large as that of ordinary 
{\it orbital} angular momentum known from Newtonian mechanics.

The electron, besides the electric charge, carries also {\it
  gravitational charges}. In the framework of general relativity, the
electron's mass is the only gravitational charge, and we expect a
mass-energy ring current inducing a gravitational moment $\mu_{\rm
  GR}$. In the Einstein-Cartan theory, which is the appropriate
gravitational theory for a Dirac field in much the same way as general
relativity, i.e. Einstein's theory, is appropriate for classical point
particles (without spin), there feature {\it two} types of
gravitational charges: mass-energy and spin. In such a theoretical
framework, we would search, already in the special-relativistic realm,
for a translational gravitational moment $M_{\rm T}$, linked to the
mass, and a Lorentz (or rotational) gravitational moment $M_{\rm L}$,
linked to the spin aspect. The latter one is to be understood as the
spin charge which is carried around by itself.

The expected units of the translational and Lorentz gravitomoments
$\mu_{\rm G_1}=\hbar/2$ and $\mu_{\rm G_2}= \hbar^2/4m_{\rm e}$ are
found by substituting in Bohr's magneton $\mu_{\rm B}$ the electric
charge by the mass or the spin of the electron, respectively. Then, in
analogy to electrodynamics, we should expect {\it gyrogravitational}
ratios for the different gravitational moments, namely $\mu_{\rm GR}/
(s_z/\hbar) =g_0\,\hbar/2$, $M_{\rm T}/(s_z/\hbar) =g_1\,\hbar/2$, and
$M_{L}/ (s_z/\hbar) =g_2\,\hbar^2/4m_{\rm e}$, with $g_0,\,g_1,\,g_2$
as the dimensionless gravitational $g$-factors. For the gravitational
$g$-factors we wouldn't be able to predict the values before going
into the corresponding computations.  But our guess, for reasons of
analogy with electrodynamics, would be $g_0=g_1=g_2=2$.

How could we thoroughly check these ideas? Again appealing to
analogies to electrodynamics, the following procedure is near at hand:
By using a Gordon type decomposition of the energy-momentum and the
spin currents in {\it special} relativity, we can provisionally define
the gravitational moments.  Then we couple the Dirac Lagrangian
minimally to the gravitational field, reshuffle the Lagrangian
suitably, and identify the factor(s) in front of the gravitational
field strength(s) as gravitational moment(s).  In general relativity,
the gravitational field strength is the Riemannian curvature, in the
Einstein-Cartan theory, both the torsion and the (Riemann-Cartan)
curvature represent the field strengths, respectively. In this article
we make the first step in identifying the gravitational moments of the
classical Dirac field within the framework of special relativity.

We find it surprising that, apart form Kobzarev and Okun \cite{KO},
nobody seems to care about the gravitational moments of the Dirac
field. It is true that the effects will be so small that there is no
hope for measuring these moments in the near future. In any case, as a
matter of principle, the gravitational moments of a fundamental
particle belong to its basic properties and will enter any unification
attempt.

We would like to dedicate this article to {\it Engelbert Sch\"ucking}
on the occasion of his 70th birthday.

\section{Dirac--Yang-Mills theory}

The Dirac Lagrangian is given by the {\em hermitian} four-form
\begin{equation}
L_{\rm D}=L(\vartheta^\alpha,\Psi,D\Psi)=
\frac{i}{2}{\hbar}\left\{\overline{\Psi}\,{}^\ast\gamma\wedge D\Psi
+\overline{D\Psi}\wedge{}^\ast\gamma\,\Psi\right\}+{}^\ast mc\,
\overline{\Psi}\Psi\,.
\label{10-4.10}
\end{equation}
The coframe $\vartheta^\alpha$ necessarily occurs in the Dirac
Lagrangian, even in special relativity. We are using the formalism of
Clifford-valued exterior forms, see the Appendix for the basic
definitions. For the mass term, we use the short-hand notation
${}^\ast m:=m\,\eta=m\,{}^\ast 1$. The hermiticity of the Lagrangian
(\ref{10-4.10}) leads to a charge current which admits the usual
probabilistic interpretation.

Here we assume a Minkowski spacetime geometry (no gravity). However,
the fermions may carry various {\it internal} charges. This is
reflected in the gauge covariant derivatives, $D=d + {\cal A}$, where
${\cal A}={e\over\hbar} A^{{}_K}\tau_{{}_K}$ is a Lie algebra-valued
one-form, the gauge potential, with a set of matrices $\tau_{{}_K}$
specifying a representation of the generators of the gauge symmetry.
The charge $e$ stands for the coupling constant of a fermion--gauge
field interaction. In the commutator
$[\tau_{{}_K},\tau_{{}_L}]=f^{{}_M}{}_{_{KL}} \tau_{{}_M}$, the
structure constants $f^{{}_M}{}_{_{KL}}$ are totally antisymmetric for
semisimple groups. The Abelian case is also trivially included, when
the internal symmetry reduces to the one-parameter group with a single
generator $\tau_1=i$ (then $f^{{}_M}{}_{_{KL}}\equiv0$). This
corresponds to electrodynamics with $e$ as the usual electric charge.

The Lagrangian (\ref{10-4.10}) yields the Dirac equation\footnote{For
  a comprehensive study of the Dirac equation, see
  Thaller \cite{thaller}.} and its adjoint which are obtained by
independent variation with respect to the spinor fields
$\overline{\Psi},\Psi$:
\begin{eqnarray}
i\hbar{}^\ast\gamma\wedge D\,\Psi + {}^\ast  mc\,\Psi &=&0,\label{10-4.11a}\\
i\hbar\overline{D\Psi}\wedge{}^\ast\gamma+{}^\ast mc\,\overline{\Psi} &=&0\,.
\label{10-4.11b}
\end{eqnarray}
The three-form 
\begin{equation}
J_{{}_K}:={\delta L_{\rm D}\over \delta A^{{}_K}}=-\,ie\,\overline{\Psi}
\tau_{{}_K}{}^\ast\gamma\Psi\label{J}
\end{equation}
is the canonical Noether current, the ``isospin'' three-form, which turns 
out to be covariantly conserved, provided the equations of motion 
(\ref{10-4.11a})-(\ref{10-4.11b}) are substituted:
\begin{equation}
DJ_{{}_K}=dJ_{{}_K} + {e\over\hbar}\,f^{{}_M}{}_{_{KL}}A^{{}_L}\wedge 
J_{{}_M}=0.\label{DJ}
\end{equation}

Using the Dirac equation, we can, following Gordon \cite{gordon},
decompose the Noether current into two pieces, the {\it convective}
and the {\it polarization} currents. The derivation is rather short:
{\it assuming the mass to be nonzero}, one can resolve
(\ref{10-4.11a})-(\ref{10-4.11b}) with respect to the spinor fields:
\begin{equation}
\Psi={i\hbar\over mc}{}^\ast({}^\ast\gamma\wedge D\Psi), \quad
\overline{\Psi}={i\hbar\over mc}{}^\ast(\overline{D\Psi}\wedge
{}^\ast\gamma).\label{psi}
\end{equation}
Now substitute this into (\ref{J}) and find
\begin{equation}
J_{{}_K}=J_{{}_K}^{(c)} + J_{{}_K}^{(p)},
\end{equation}
where the convective three-form current reads
\begin{equation}
J_{{}_K}^{(c)}:={e\hbar\over 2mc}\,{}^\ast(\overline{\Psi}\tau_{{}_K} D\Psi -
\overline{D\Psi}\tau_{{}_K}\Psi),\label{Jconv}
\end{equation}
and the polarization current is constructed as a covariant exterior 
differential of the {\it polarization} two--form $P_{{}_K}$:
\begin{equation}
J_{{}_K}^{(p)}:=DP_{{}_K}, \qquad P_{{}_K}:=-\,i\,{e\hbar\over 2mc}\,
\overline{\Psi}\tau_{{}_K}{}^\ast\widehat{\sigma}\Psi.\label{Jpol}
\end{equation}
The covariant exterior differential $D$ is defined as in (\ref{DJ}).

Unlike the total current, its convective and polarization parts are
{\it not} conserved separately in non-Abelian gauge theory. For
example, we can immediately check that
\begin{equation}
DJ_{{}_K}^{(p)}=DDP_{{}_K}={e\over\hbar}\,f^{{}_N}{}_{{}_{KL}}F^{{}_L}\wedge 
P_{{}_N},
\end{equation}
where 
\begin{equation}
F^{{}_K}:=dA^{{}_K}+{e\over2\hbar}\,f^{{}_K}{}_{_{MN}}A^{{}_M}\wedge A^{{}_N}
\end{equation}
is the two-form of the gauge field strength. [We can also write the gauge 
field strength in the form ${\cal F}:=d{\cal A} + {\cal A}\wedge{\cal A}=
{e\over\hbar}F^{{}_K}\tau_{{}_K}$. Note that $DD\Psi={\cal F}\Psi$.]
Only in the {\it Abelian} case one finds Gordon's separate 
conservation of the two currents. 

In order to understand the physical meaning of these currents, it is 
instructive to study the non-relativistic approximation for fermions in a 
weak external gauge field. In the non-relativistic approximation, a Dirac 
four-spinor is represented by a pair of two-component spinors, 
$\Psi=\left(\!\begin{array}{r}\psi\\ \chi\end{array}\!\right)$, with 
$\chi={\cal O}({v\over c})\,\psi$ (for positive energy solutions). The 
polarization two-form is then:
\begin{equation} 
P_{{}_K}={e\hbar\over m}\,dt\wedge S_{{}_K} + 
{\cal O}\left({v\over c}\right),
\qquad S_{{}_K}:=\psi^\dagger(-i\tau_{{}_K})\hbox{$\scriptstyle{1\over 2}$} 
\hbox{\boldmath $\sigma$}\psi\,d\hbox{\boldmath $x$}.\label{NonRel1}
\end{equation}
Consider the interaction term in the Lagrangian, $A^{{}_K}\wedge J_{{}_K}$.
It is clear that the convective contribution describes a (non-Abelian 
generalization of the) usual Schr\"odinger current, 
\begin{equation}
J_{{}_K}^{(c)}\approx {e\hbar\over 2mc}{}^\ast[\psi^\dagger\tau_{{}_K}d\psi 
-(d\psi^\dagger)\tau_{{}_K}\psi],
\end{equation}
whereas the polarization contribution reads
\begin{equation}
A^{{}_K}\wedge J_{{}_K}^{(p)}\approx F^{{}_K}\wedge P_{{}_K}\approx
{e\hbar\over m}\,dt\wedge S_{{}_K}\wedge F^{{}_K}.
\end{equation}
In the Abelian case (for $\tau_{{}_K}=i$), the standard result of the
Dirac electron theory is recovered, $A\wedge
J^{(p)}\approx\hbox{\boldmath $\mu\,B$} \,dt\wedge{}^{(3)}\eta$. Here
${}^{(3)}\eta$ is the volume form of a spatial hypersurface,
$\hbox{\boldmath $B$}$ is the magnetic field strength, and
$\hbox{\boldmath $\mu$}={e\hbar\over 2m}\psi^\dagger\hbox{\boldmath
  $\sigma$}\psi$.

Technically, we can derive the two currents as follows. Substituting 
(\ref{psi}) back into the Dirac equation (\ref{10-4.11a})-(\ref{10-4.11b}), 
we find the {\it squared} equation:
\begin{equation}
\left[D{}^\ast\!D - i\,{}^\ast\widehat{\sigma}\wedge{\cal F} + {}^\ast\!
\left({mc/\hbar}\right)^2\right]\Psi=0.\label{sqdir}
\end{equation}
Eq.(\ref{sqdir}) can be derived from the Lagrange form $L_{{\rm
    D}^2}=L^{(c)} + L^{(p)}$, with
\begin{eqnarray}
L^{(c)}&:=&{1\over 2}\left({\hbar^2\over mc}\,{}^\ast\overline{D\Psi}\wedge 
D\Psi + {}^\ast mc\,\overline{\Psi}\,\Psi\right),\label{Lconv}\\
L^{(p)}&:=& P_{{}_K}\wedge F^{{}_K}.\label{Lpol}
\end{eqnarray}
Since both pieces of the Lagrangian are separately gauge invariant, we 
straightforwardly recover the convective and polarization currents, 
(\ref{Jconv}) and (\ref{Jpol}), as the respective Noether currents:
\begin{equation}
J_{{}_K}^{(c)}={\delta L^{(c)}\over \delta A^{{}_K}},\qquad
J_{{}_K}^{(p)}={\delta L^{(p)}\over \delta A^{{}_K}}.
\end{equation}

\section{Gordon decomposition of energy-momentum and spin currents}

The {\it canonical} energy-momentum and spin three-forms are defined in
the standard way:
\begin{eqnarray}
\Sigma_\alpha &:=& e_\alpha\rfloor L - {\partial L\over \partial D\Psi}\,
(e_\alpha\rfloor D\Psi) -(e_\alpha\rfloor\overline{D\Psi})\, {\partial L\over 
\partial\overline{D\Psi}}\,,\label{sigma}\\
\tau_{\alpha\beta}&:=&{\partial L\over \partial D\Psi}\,
\ell_{\alpha\beta}\Psi + \overline{\Psi}\ell_{\alpha\beta}
{\partial L\over\partial\overline{D\Psi}}\,.\label{tau}
\end{eqnarray}
Recalling that the spinor generators of the Lorentz group are 
$\ell_{\alpha\beta}={\frac i 4}\,\widehat{\sigma}_{\alpha\beta}$, we find
for the Dirac Lagrangian (\ref{10-4.10}), with $L=L_{\rm D}$ inserted into 
(\ref{sigma})-(\ref{tau}),
\begin{eqnarray}
\Sigma_\alpha &=& {i\hbar\over 2}\left(\overline{\Psi}\,{}^\ast\gamma
D_\alpha\Psi - D_\alpha\overline{\Psi}\,{}^\ast\gamma\Psi\right),\label{sigmaD}\\
\tau_{\alpha\beta}&=& {\hbar\over 4}\,\vartheta_\alpha\wedge\vartheta_\beta
\wedge\overline{\Psi}\gamma\gamma_5\Psi.\label{tauD}
\end{eqnarray}
Here we denoted $D_\alpha:=e_\alpha\rfloor D$ and took into account
that $L_{\rm D}\cong 0$ (upon substitution of the field equations).

Translational and Lorentz invariance of the Dirac Lagrangian yield the
well-known conservation laws for energy-momentum and angular momentum:
\begin{equation}
D\Sigma_\alpha=0,\qquad D\tau_{\alpha\beta} + \vartheta_{[\alpha}\wedge
\Sigma_{\beta]}=0.\label{cons1}
\end{equation}
Here $D$ denotes the Lorentz covariant exterior derivative containing
the Levi-Civita connection $\Gamma_\alpha{}^\beta$ of flat
Minkowski space. Thus we have, e.g., $D\Sigma_\alpha=d\Sigma_\alpha-
\Gamma_\alpha{}{}^\beta\wedge\Sigma_\beta$.

Let us perform the Gordon type decomposition of the energy-momentum
current \cite{heyde,erice79,Au,seitz1,seitz2}. At first, we substitute
(\ref{psi}) into (\ref{sigmaD}) and obtain:
\begin{eqnarray}
\Sigma_\alpha&=&{\hbar^2\over 2 mc}\Big[{}^\ast(\overline{D\Psi})D_\alpha\Psi
+ D_\alpha\overline{\Psi}{}^\ast D\Psi \nonumber\\
&&+ i\left(\overline{D\Psi}\wedge{}^\ast
\widehat{\sigma}D_\alpha\Psi - D_\alpha\overline{\Psi}{}^\ast\widehat{\sigma}
\wedge D\Psi\right)\Big].\label{sigD1}
\end{eqnarray}
The Dirac equation (\ref{10-4.11a})-(\ref{10-4.11b}) yields:
\begin{eqnarray}
i\hbar{}^\ast\gamma D_\alpha\,\Psi &=& i\hbar(e_\alpha\rfloor{}^\ast
\gamma)\wedge D\Psi + \eta_\alpha mc\,\Psi,\label{dir1a}\\
i\hbar D_\alpha\overline{\Psi}{}^\ast\gamma &=& i\hbar\overline{D\Psi}
\wedge(e_\alpha\rfloor{}^\ast\gamma) - \eta_\alpha mc\,\overline{\Psi}.
\label{dir1b}
\end{eqnarray}
Substituting this into (\ref{sigmaD}), we find another representation
for the energy-mo\-men\-tum, provided we use (\ref{psi}) again:
\begin{eqnarray}
\Sigma_\alpha&=& {i\hbar\over 2}\left((e_\alpha\rfloor\overline{\Psi}{}^\ast
\gamma)\wedge D\Psi - \overline{D\Psi}\wedge(e_\alpha\rfloor{}^\ast\gamma\Psi)
\right) + \eta_\alpha mc\,\overline{\Psi}\Psi\nonumber\\
&=& {\hbar^2\over 2 mc}\Big[(e_\alpha\rfloor{}^\ast\overline{D\Psi})\wedge
D\Psi + \overline{D\Psi}\wedge(e_\alpha\rfloor{}^\ast D\Psi)\nonumber\\
&& +i\left(D_\alpha\overline{\Psi}{}^\ast\widehat{\sigma}\wedge D\Psi -
\overline{D\Psi}\wedge{}^\ast\widehat{\sigma}D_\alpha\Psi\right)\nonumber\\
&&-2i\,\overline{D\Psi}\wedge(e_\alpha\rfloor{}^\ast\widehat{\sigma})\wedge
D\Psi\Big] + \eta_\alpha mc\,\overline{\Psi}\Psi.\label{sigD2}
\end{eqnarray}
Then, the sum of (\ref{sigD1}) and (\ref{sigD2}) yields
\begin{equation}
\Sigma_\alpha = \Sigma_\alpha^{(c)} + \Sigma_\alpha^{(p)},\label{decmom}
\end{equation}
where 
\begin{eqnarray}
\Sigma_\alpha^{(c)}&:=&{mc\over 2}\,\overline{\Psi}\Psi\,\eta_\alpha
+ {\hbar^2\over 4mc}\Big[{}^\ast(\overline{D\Psi})D_\alpha\Psi
+ D_\alpha\overline{\Psi}{}^\ast D\Psi \nonumber\\
&&+ (e_\alpha\rfloor{}^\ast\overline{D\Psi})\wedge D\Psi + \overline{D\Psi}
\wedge(e_\alpha\rfloor{}^\ast D\Psi)\Big],\label{momconv}\\
\Sigma_\alpha^{(p)}&:=& D{\hbox{\it \v{M}}}_\alpha,\label{mompol}\\
{\hbox{\it \v{M}}}_\alpha &:=& -\,{i\hbar^2\over 4mc}\,[\overline{\Psi}\,
(e_\alpha\rfloor{}^\ast\,\hat{\sigma})\wedge D\Psi + \overline{D\Psi}
\wedge(e_\alpha\rfloor{}^\ast\,\hat{\sigma})\,\Psi].\label{Mtil}
\end{eqnarray}

Since we are in flat Minkowski spacetime, the last term in
(\ref{decmom}), the {\it polarization} part of the energy-momentum
current, is identically conserved,
\begin{equation}
D\Sigma_\alpha^{(p)}= DD{\hbox{\it \v{M}}}_\alpha =0.\label{cons2}
\end{equation}
Recalling (\ref{cons1}), we immediately obtain a separate conservation
of the first term on the right hand side of (\ref{decmom}), which we
naturally call the {\it convective energy-momentum} current:
\begin{equation}
D\Sigma_\alpha^{(c)}=0.
\end{equation}
Moreover, one can immediately check that the convective current is
symmetric,
\begin{equation}
\vartheta_{[\alpha}\wedge\Sigma_{\beta]}^{(c)}=0.\label{convsym}
\end{equation}

It is remarkable to observe that, similar as the convective current
(\ref{Jconv}) represents a Noether current for the convective
Lagrangian (\ref{Lconv}), the convective energy-momentum
(\ref{momconv}) precisely turns out to be the canonical Noether
current (\ref{sigma}) for $L=L^{(c)}$. The proof is straightforward:
substitute (\ref{Lconv}) into (\ref{sigma}) and compare with
(\ref{momconv}).

As we saw, the ordinary polarization current (\ref{Jpol}) also emerges
as a Noether current for the ``Pauli-type'' polarization Lagrangian
(\ref{Lpol}) which describes the interaction of the {\it moment
  two-form} with the background gauge field strength $F^{{}_K}$. A
guess would be that the polarization energy-momentum should arise in a
similar way as a canonical current {}from the respective
``Pauli-type'' polarization Lagrangian when the gravitational field is
``switched on''. The discussion of the precise form of this Lagrangian
(as well as of the nature of the gravitational field represented by
the coframe) will be considered in a separate publication, see also
\cite{fwhni,bad90,He20,AHL,lemke,MMMa96}.

Here, however, we have to complete our derivations and to consider the Gordon
type decomposition of the spin current. We have good reasons to expect a 
similar structure of the decomposed three-form $\tau_{\alpha\beta}$. At first,
we note that the canonical spin current (\ref{tau}) for the {\it convective}
Lagrangian (\ref{Lconv}) reads:
\begin{eqnarray}
\tau_{\alpha\beta}^{(c)}&=&{\partial L^{(c)}\over \partial D\Psi}\,
\ell_{\alpha\beta}\Psi + \overline{\Psi}\ell_{\alpha\beta}
{\partial L^{(c)}\over\partial\overline{D\Psi}}\nonumber\\
&=&-\,{i\hbar^2\over 8mc}\left({}^\ast\overline{D\Psi}
\widehat{\sigma}_{\alpha\beta}\Psi - \overline{\Psi}\widehat{\sigma}
_{\alpha\beta}{}^\ast D\Psi\right).\label{tauC}
\end{eqnarray}
Now we use the standard trick by substituting (\ref{psi}) into the 
spin current (\ref{tauD}):
\begin{eqnarray}
\tau_{\alpha\beta}&=& {\hbar\over 4}\,\vartheta_\alpha\wedge\vartheta_\beta
\wedge\overline{\Psi}\gamma\gamma_5\Psi = {\hbar\over 8}\,\overline{\Psi}
({}^\ast\gamma\widehat{\sigma}_{\alpha\beta} + \widehat{\sigma}_{\alpha\beta}
{}^\ast\gamma)\Psi \nonumber\\
&=& {i\hbar^2\over 8mc}\Big(-{}^\ast\overline{D\Psi}
\widehat{\sigma}_{\alpha\beta}\Psi + \overline{\Psi}\widehat{\sigma}
_{\alpha\beta}{}^\ast D\Psi\nonumber\\
&&-i\overline{D\Psi}\wedge{}^\ast\widehat{\sigma}\widehat{\sigma}_{\alpha\beta}
\Psi - i\overline{\Psi}\widehat{\sigma}_{\alpha\beta}{}^\ast\widehat{\sigma}
\wedge D\Psi\Big).\label{tauD1}
\end{eqnarray}
We immediately recognize the second line as the {\it convective spin} 
(\ref{tauC}). The last line should thus be related to the polarization spin
current. The identities (\ref{sisi3})-(\ref{sisi4}) (see the Appendix) play 
the crucial role here. Substituting them into (\ref{tauD1}), we obtain 
straightforwardly,
\begin{equation}
\tau_{\alpha\beta} = \tau_{\alpha\beta}^{(c)} + DM_{\alpha\beta} +
\vartheta_{[\alpha}\wedge {\hbox{\it \v{M}}}_{\beta]},\label{decspin}
\end{equation}
where we define the {\it Lorentz gravitational moment} two-form by
\begin{eqnarray}
  M_{\alpha\beta}&:=&{\hbar^2\over
    16mc}\overline{\Psi}({}^\ast\widehat{\sigma}
  \widehat{\sigma}_{\alpha\beta} +
  \widehat{\sigma}_{\alpha\beta}{}^\ast \widehat{\sigma})\Psi
  \label{lorgrav}\\ &=& {\hbar^2\over
    8mc}\left(\overline{\Psi}\Psi\,\eta_{\alpha\beta} -
    i\,\overline{\Psi}\gamma_5\Psi\,\vartheta_\alpha\wedge
    \vartheta_\beta\right),
\label{lorgrav1}
\end{eqnarray}
whereas ${\hbox{\it \v{M}}}_\alpha$ is given in (\ref{Mtil}). The
Lorentz moment $M_{\alpha\beta}$ is very simple in structure. It is
additively built up from a scalar and a pseudoscalar piece, i.e., from
its $36$ components only $2$ are independent.

Substituting the Gordon decompositions (\ref{decmom}) and
(\ref{decspin}) into the conservation law of angular momentum
(\ref{cons1}), we find the {\it separate} conservation of the
convective spin,
\begin{equation}
D\tau_{\alpha\beta}^{(c)}=0.
\end{equation}

\section{Relocalization of energy-momentum and spin}

Like for internal symmetries, also in the case of the Poincar\'e group,
the Noether currents are only determined up to an exact two-form. This
non-uniqueness has troubled physicists already for quite some time.
Within gravitational theory, the question of the ``correct''
energy-momentum current of matter is as old as general relativity
itself \cite{Hilbert,EinsteinHamilton}, see also the review
\cite{HehlRoMP}.  But only Belinfante \cite{Belinfante}, in the
framework of special relativity, and Rosenfeld \cite{Rosenfeld},
within general relativity, gave a general prescription of how one can
find the metric or {\it Hilbert} energy-momentum current {}from the
canonical or {\it Noether} energy-momentum current of an arbitrary
matter field $\Psi$. The Hilbert current acts as source on the right
hand side of the Einstein field equation, whereas the Noether current
is of central importance in special-relativistic canonical field
theory. We will now turn our attention to this interrelationship
between the different energy-momentum currents within the framework of
special relativity.

The Noether law for energy-momentum in (\ref{cons1}) also holds for an
energy-momentum current which is supplemented by a $D$-exact 
form:
\begin{equation}\label{supplement}
\hat\Sigma_\alpha( X ):=
\Sigma_\alpha-D X _\alpha .
\end{equation}
In special relativity, $DD=0$. This is the only property of $D$ that
is needed in this context.  The $X_\alpha$ does not interfere with the
Noether law:
\begin{equation}\label{sub1}
D\Sigma_\alpha=D\hat\Sigma_\alpha+DD X _\alpha=D\hat\Sigma_\alpha=0\,.
\end{equation}
If we insert $\Sigma_\alpha=\hat\Sigma_\alpha( X )+D X _\alpha$ into
the left hand side of the Noether law for angular momentum in
(\ref{cons1}), we find
\begin{equation}\label{sub2}
D\tau_{\alpha\beta}+\vartheta_{[\alpha}\wedge\Sigma_{\beta]}
  =D(\tau_{\alpha\beta}-\vartheta_{[\alpha}\wedge X _{\beta]})
  +\vartheta_{[\alpha}\wedge\hat\Sigma_{\beta]}\,.
\end{equation}
If a relocalized spin $\hat\tau_{\alpha\beta}$ is required to fulfill
again a law of the type given in (\ref{cons1}), i.e.
$D\hat\tau_{\alpha\beta}+\vartheta_{[\alpha}\wedge\hat\Sigma_{\beta]}=0$,
then
\begin{equation}\label{sub3}
  \hat\tau_{\alpha\beta}( X ,Y):=\tau_{\alpha\beta}
  -\vartheta_{[\alpha}\wedge X _{\beta]}-DY_{\alpha\beta}\,,
\end{equation}
where $DY_{\alpha\beta}$ is an additional $D$-exact form with
$Y_{\alpha\beta}=-Y_{\beta\alpha}$. Thus a relocalization of the
energy-momentum is, up to a $D$-exact form, accompanied by an induced
transformation of the canonical spin. Therefore we have the following
result: \medskip

{\it The canonical currents $(\Sigma_\alpha,\tau_{\alpha\beta})$
  fulfill the Noether laws (\ref{cons1}). Take arbitrary two-forms $ X
  _\alpha$ and $Y_{\alpha\beta}=-Y_{\beta\alpha}$ as superpotentials.
  Then the relocalized currents
\begin{eqnarray}
\Sigma_\alpha &\rightarrow & \hat\Sigma_\alpha( X )
  =\Sigma_\alpha-D X _\alpha\,,\label{reloc}\\
  \tau_{\alpha\beta}&\rightarrow &\hat\tau_{\alpha\beta}( X ,Y)
  =\tau_{\alpha\beta}-\vartheta_{[\alpha}\wedge X _{\beta]}
  -DY_{\alpha\beta}\,,\label{reloc1}
\end{eqnarray}
satisfy the same relations}
\begin{equation}\label{cons10}
  D\hat\Sigma_\alpha=0\,,\qquad
  D\hat\tau_{\alpha\beta}+\vartheta_{[\alpha}\wedge\hat\Sigma_{\beta]}=0\,.
\end{equation}

Accordingly, the Noether identities turn out to be invariant under the
{\it relocalization transformation} (\ref{reloc})-(\ref{reloc1}).
As a consequence, the total energy-momentum $P_\alpha$ and the total
angular momentum $J_{\alpha\beta}$, up to boundary terms, remain
invariant under (\ref{reloc})-(\ref{reloc1}),
\begin{equation}
  \hat P_\alpha\stareq P_\alpha-\int_{\partial H_t} X _\alpha\,\qquad
  \hat J_{\alpha\beta}\stareq J_{\alpha\beta}-\int_{\partial H_t}
  (x_{[\alpha}\wedge X _{\beta]}+Y_{\alpha\beta})\,,
\end{equation}
where $H_t$ denotes a timelike hypersurface in Minkowski space and
$\partial H_t$ its two-dimensional boundary. Provided the
superpotentials $ X _\alpha$ and $Y_{\alpha\beta}$ approach zero at
spacelike asymptotic infinity sufficiently fast, the total quantities
are not affected by the relocalization procedure.  \medskip

Let us put our results of the {\it Gordon decomposition} in the last
section into this general framework. If we choose as superpotentials
the respective gravitational moments,
\begin{equation}
   X _\alpha={\hbox{\it \v{M}}}_\alpha\,,\qquad
  Y_{\alpha\beta}=M_{\alpha\beta}\,,\label{XYMM}
\end{equation}
then the relocalized currents turn out to be the {\it convective} pieces:
\begin{eqnarray}\label{gord1}
  \Sigma_\alpha^{(c)} &=&\Sigma_\alpha-D\check M_\alpha
  =\Sigma_\alpha-\Sigma_\alpha^{(p)}\,,\\ \tau_{\alpha\beta}^{(c)}
  &=&\tau_{\alpha\beta} -\vartheta_{[\alpha}\wedge\check
  M_{\beta]}-DM_{\alpha\beta}
  =\tau_{\alpha\beta}-\tau_{\alpha\beta}^{(p)}\,.\label{gord2}
\end{eqnarray}
{}From explicit calculations we know, see (\ref{convsym}), that the
convective current (\ref{gord1}) is symmetric, as one would expect
for a Schr\"odinger type energy-momentum current. Consequently, in
special relativity, the decomposed currents have the following
properties:
\begin{equation}\label{gord3}
D\Sigma_\alpha^{(c)}= 0\,,\qquad 
  D\Sigma_\alpha^{(p)}= 0\,,\qquad
  \vartheta_{[\alpha}\wedge\Sigma_{\beta]}^{(c)}= 0\,,
\end{equation}
\begin{equation}
  D\tau_{\alpha\beta}^{(c)}= 0\,,\qquad
  D\tau_{\alpha\beta}^{(p)}+\vartheta_{[\alpha}\wedge
  \Sigma_{\beta]}^{(p)}= 0 \,.
\end{equation} 
Thus a Gordon decomposition in Minkowski spacetime is nothing else but
a {\it specific relocalization} of the currents. It yields a symmetric
energy-momentum current $\Sigma_\alpha^{(c)}$ with a nonvanishing {\it
  conserved} spin current $\tau_{\alpha\beta}^{(c)}$. The spin tensor
of Hilgevoord et al.\ \cite{hilge1,hilge2}, which was constructed
outside of the framework of Lagrangian formalism, coincides with our
convective spin current\footnote{\dots up to a factor of $2$
due to different conventions.} $\tau_{\alpha\beta}^{(c)}$. 

\section{Trivial Lagrangians and relocalization}

For the purpose of understanding relocalization from a Lagrangian
point of view, let us consider an arbitrary {\it three}--form
$U(\Psi,D\Psi)$ constructed from a matter field $\Psi$ and its
derivatives. Here we discard other possible arguments of $U$ (like the
coframe, e.g.). In general, $\Psi$ can be any matter field, not
necessarily the Dirac four-spinor.

If the form $U$ is invariant under spacetime translations (coordinate
transformations) and Lorentz transformations, then, using the standard
Lagrange-Noether machinery, cf. \cite{PR}, one derives the first and
the second Noether {\it identities}:
\begin{eqnarray}
D{\buildrel \circ\over X}_\alpha &\equiv& -\,e_\alpha\rfloor d\,U + 
(e_\alpha\rfloor D\Psi)\,{\delta U\over \delta\Psi},\label{noe1}\\
D{\buildrel \circ\over Y}_{\alpha\beta} + \vartheta_{[\alpha}\wedge
{\buildrel \circ\over X}_{\beta]}&\equiv& -\,\ell_{\alpha\beta}\Psi\,
{\delta U\over \delta\Psi}.\label{noe2}
\end{eqnarray}
Here the {\it two}-forms
\begin{eqnarray}
  {\buildrel \circ\over X}_\alpha &:=& e_\alpha\rfloor U -
  (e_\alpha\rfloor D\Psi)\,{\partial U\over \partial D\Psi}
  ,\label{sig0}\\ {\buildrel \circ\over
    Y}_{\alpha\beta}&:=&\ell_{\alpha\beta}\Psi\, {\partial U\over
    \partial D\Psi}\label{tau0},
\end{eqnarray}
are analogs of the three-forms of canonical energy-momentum and spin
derived from the ``Lagrangian'' $U$. We use the standard notation
\begin{equation}
{\delta U\over \delta\Psi}:={\partial U\over \partial\Psi} -
D{\partial U\over \partial D\Psi}.
\end{equation}
We will {\it not} assume, however, that the matter field $\Psi$
satisfies the ``equation of motion'' ${\delta U\over \delta\Psi}=0$.
The relations (\ref{noe1}) and (\ref{noe2}) are thus {\it strong
  identities} valid for all matter field configurations. The first
term on the right-hand side of (\ref{noe1}) is usually absent in the
first Noether identity due to the fact that the Lagrangian is a form
of maximal rank (i.e.\ four in standard spacetime). Here we have a
three-form $U$ in four-dimensional spacetime, and the four-form $d\,U$
is, in general, nontrivial.

Let us now treat the form $L=d\,U$ as a specific Lagrangian. It is
straightforward to find for the variation:
\begin{equation}
\delta U = \delta\Psi\,{\partial U\over \partial\Psi} + 
\delta D\Psi\wedge{\partial U\over \partial D\Psi} =
\delta\Psi\,{\delta U\over \delta\Psi} + d\left(\delta\Psi\,
{\partial U\over \partial D\Psi}\right).
\end{equation}
Hence, for our specific Lagrangian, we have
\begin{equation}
\delta L = d(\delta U)= \delta\Psi\,D\left({\delta U\over \delta\Psi}\right)
+ \delta (D\Psi)\wedge{\delta U\over \delta\Psi}.
\end{equation}
For the partial derivatives, this yields 
\begin{eqnarray}
{\partial L\over \partial\Psi}&=&D\left({\delta U\over \delta\Psi}\right),\\
{\partial L\over \partial D\Psi}&=&{\delta U\over \delta\Psi}.\label{dLdpsi}
\end{eqnarray}
Consequently, for the Lagrangian $L=d\,U$, the equation of motion
\begin{equation}
{\partial L\over \partial\Psi}= {\partial L\over \partial\Psi} -
D{\partial L\over \partial D\Psi}\equiv 0
\end{equation}
is identically satisfied. Actually, this is no surprise: it is well
known that a Lagrangian, which is a total differential, has trivial
dynamics.

Nevertheless, although the dynamics is trivial, the conserved currents
are {\it not} trivial. In particular, the canonical energy-momentum
and spin three-forms are defined as usual by
\begin{eqnarray}
  \Sigma_\alpha &=& e_\alpha\rfloor L - {\partial L\over \partial
    D\Psi}\, (e_\alpha\rfloor D\Psi),\\ \tau_{\alpha\beta}&=&{\partial
    L\over \partial D\Psi}\, \ell_{\alpha\beta}\Psi .
\end{eqnarray}
And now we are approaching the crucial point. We recall that $L=d\,U$.
Thus we can substitute (\ref{dLdpsi}) into these currents. Then, for
the corresponding energy-momentum, we immediately find
\begin{equation}
  \Sigma_\alpha = e_\alpha\rfloor d\,U - (e_\alpha\rfloor D\Psi)\,
  {\delta U\over \delta\Psi} = - D{\buildrel \circ\over X}_\alpha,
\end{equation}
where we made use of the first Noether identity (\ref{noe1}).
Analogously, for the spin current, we have 
\begin{equation}
  \tau_{\alpha\beta}=\ell_{\alpha\beta}\Psi\,{\delta U\over
    \delta\Psi} = - D{\buildrel \circ\over Y}_{\alpha\beta} -
  \vartheta_{[\alpha}\wedge {\buildrel \circ\over X}_{\beta]},
\end{equation}
where we used the second Noether identity (\ref{noe2}).

This observation underlies the relocalization described above of
energy-mo\-men\-tum and spin. Indeed, our derivation shows that if one
adds to any matter field Lagrangian $L_{\Psi}$ a total
divergence\footnote{In previous papers, see \cite{MMM96}, this
  Lagrangian prescription was used ``on shell" to generate the
  transition to chiral fermions. However, in the massless limit these
  fields carry no spin but rather helicity.} $d\,U$, then, for the new
Lagrangian $L_{\Psi} + d\,U$, the canonical energy-momentum and spin
currents are relocalized according to (\ref{reloc})-(\ref{reloc1}),
with the superpotentials
\begin{equation}
X_\alpha={\buildrel \circ\over X}_\alpha,\qquad
Y_{\alpha\beta}={\buildrel \circ\over Y}_{\alpha\beta}.\label{XY0}
\end{equation}
In this sense, one can say that a relocalization is {\it generated} by
the three-form $U$ via (\ref{sig0})-(\ref{tau0}). 

Again looking back to the Gordon decomposition of energy-mo\-men\-tum and
spin as a special case of the relocalization procedure, we are now able to 
{\it generate} the corresponding results by means of the simple 
{\it three-form} 
\begin{equation}\label{generating}
  U={1\over 2}\,\vartheta^\alpha\wedge\check M_\alpha\,=-\,
  {i\hbar^2\over 4mc}\left(\overline{\Psi}\,^\ast\widehat{\sigma}\wedge D\Psi
    - \overline{D\Psi}\wedge\,^\ast\widehat{\sigma}\Psi \right).\label{Udir}
\end{equation}
If we substitute it into (\ref{sig0})-(\ref{tau0}), we find, indeed,
\begin{equation}\label{xcircle}
{\buildrel \circ\over X}_\alpha=\check M_\alpha\,,\qquad {\rm  and}\qquad 
{\buildrel \circ\over Y}_{\alpha\beta}=M_{\alpha\beta}\,,
\end{equation}
compare with (\ref{XYMM}) and (\ref{XY0}). It is remarkable that the 
{\it translational} moment $\check M_\alpha$, via $U$, also generates the
corresponding {\it Lorentz} moment $M_{\alpha\beta}$.

Moreover, the {\it same} three-form $U$ generates the relocalization
of the isospin current $J_{{}_K}$,
\begin{equation}
J_{{}_K}\rightarrow J_{{}_K} - D{\buildrel \circ\over Z}_{{}_K},
\end{equation}
where 
\begin{equation}
{\buildrel \circ\over Z}_{{}_K}={e\over\hbar}\,\tau_{{}_K}\Psi\,
{\partial U\over\partial D\Psi}.\label{Z0}
\end{equation}
The proof goes along the same lines as above. We just formulate the
relevant Noether identity which arises from the invariance of the
three-form $U$ with respect to the gauge transformations under
consideration. Substituting (\ref{Udir}) into (\ref{Z0}), we recover
the polarization moment two-form
\begin{equation}
{\buildrel \circ\over Z}_{{}_K}=P_{{}_K}=-\,i\,{e\hbar\over 2mc}\,
\overline{\Psi}\tau_{{}_K}{}^\ast\widehat{\sigma}\Psi.
\end{equation}

\section{Belinfante symmetrization of the\\ energy-momentum current}

A simple way, within special relativity, to arrive at the ``generic''
symmetric energy-momentum current, i.e., at the Hilbert current, is to
require that the {\it relocalized spin current vanishes}. This is what
the Belinfante-Rosenfeld symmetrization amounts to. Therefore the
Belinfante-Rosenfeld energy-momentum current $ t _\alpha$ can be
defined as
\begin{equation} 
t _\alpha:= \hat\Sigma_\alpha( X )\qquad 
  {\rm with}\qquad \hat\tau_{\alpha\beta}( X ,Y)= 0\,.
\end{equation}
The last equation, together with (\ref{reloc1}), yields
$\tau_{\alpha\beta}=\vartheta_{[\alpha}\wedge X _{\beta]}+DY_{\alpha\beta}$
which can be resolved with respect to the superpotential $ X ^\beta$
as follows:
\begin{equation}\label{sup1}
 X^\beta={\mu}{}^\beta
  -2e_\gamma\rfloor DY^{\gamma\beta}-{1\over 2}\vartheta^\beta\wedge
  \bigl(e_\gamma\rfloor e_\delta\rfloor DY^{\gamma\delta}\bigr)\,.
\end{equation} 
Here
\begin{equation}\label{sup2}
{\mu}{}^\beta:= 
  2e_\gamma\rfloor\tau^{\gamma\beta} +{1\over 2}\vartheta^\beta\wedge
  \bigl(e_\gamma\rfloor e_\delta\rfloor\tau^{\gamma\delta}\bigr)
\end{equation}
is the {\it spin energy potential}. Then the first Noether law
in (\ref{cons10}) reads alternatively $D t _\alpha=0$.\medskip

Let us collect the key formulae for our Belinfante-Rosenfeld current
with the {\it specific} superpotential $X_\alpha$ of
(\ref{sup1})-(\ref{sup2}):
\begin{equation}
t _\alpha=\Sigma_\alpha-D X _\alpha\,,\qquad
  \vartheta_{[\alpha}\wedge t _{\beta]}= 0\,,\qquad D t
  _\alpha=0\,.
\end{equation} 
For $Y_{\alpha\beta}=0$, these are the
familiar Belinfante-Rosenfeld relations \cite{Belinfante,Rosenfeld}.
For particles with spin zero, the improved energy-momentum current can
be derived by a suitable choice of the superpotential
$Y_{\alpha\beta}$, cf.\ \cite{He20}.

It is remarkable that, for a matter field of any spin, we can find a
relocalized Belinfante-Rosenfeld energy-momentum current $t _\alpha$,
with $D t _\alpha=0$. If we consider the motion of a ``test'' field in
a Minkowski spacetime, then our procedure shows that we can always
attach to this motion a geodesic line, irrespective of the spin. 

As we saw in the previous section, a relocalization of energy-momentum
and spin can be generated by a superpotential three-form $U$. However,
the finding of an explicit $U$ for the Belinfante relocalization turns
out to be a non-trivial problem. Although the general prescription
(\ref{sup1}) involves $Y_{\alpha\beta}$, a symmetrization of the
energy-momentum is already achieved for $Y_{\alpha\beta}=0$.
Moreover, our discussion here was confined to flat Minkowski geometry;
but in Riemannian spacetime (which is the arena of general relativity
theory) the Belinfante relocalization necessarily demands
$Y_{\alpha\beta}=0$, as was shown in \cite{RCbel}.

Accordingly, a puzzling feature of the Belinfante relocalization for
the Dirac energy-momentum is the apparent impossibility of
constructing a generating form $U$, with $Y_{\alpha\beta}=0$. Indeed,
for the Dirac field, the spin current is given by (\ref{tauD}). Hence
the spin energy potential (\ref{sup2}) reads:
\begin{equation}
  \mu_\alpha={\hbar\over
    4}\vartheta_\alpha\wedge\overline{\Psi}\gamma\gamma_5
  \Psi.\label{muD}
\end{equation}
{}From (\ref{tau0}) it is clear that $Y_{\alpha\beta}=0$ if and only if $U$ 
does not depend on the differentials $D\Psi$, i.e. $\partial U/\partial D\Psi
=0$. Then (\ref{sig0}) and (\ref{sup1}) yield
\begin{equation}
\mu_\alpha=e_\alpha\rfloor U.
\end{equation}
Contracting with the coframe $\vartheta^\alpha$, we find
$\vartheta^\alpha \wedge\mu_\alpha= 3U$. However, for the Dirac spin
energy potential (\ref{muD}) we get
$\vartheta^\alpha\wedge\mu_\alpha\equiv 0$. Therefore we have to
conclude that there is no such three-form $U$ which can generate the
Belinfante relocalization in the Dirac theory --- provided one starts
with the canonical currents (\ref{sigmaD})-(\ref{tauD}).  Probably the
latter requirement has to be given up. One could start with the
convective currents (\ref{momconv}) and (\ref{tauC}) as well. But we
will leave that for future consideration.

\section{Properties of the gravitational moments and non-relativistic limit}

Let us find out the dimensions of the gravitational moments. Recall that the 
Dirac fields has dimension $[\Psi]=[\overline{\Psi}]=length^{-3/2}$, whereas 
$[\vartheta^\alpha]=length$, and $[e_\alpha]=length^{-1}$. Thus, we have for 
the two-forms $[\widehat{\sigma}]=[{}^\ast\widehat{\sigma}]=length^2$, and
we immediately get
\begin{equation}
[\check M_\alpha]=[mc], \qquad [M_{\alpha\beta}]=[\hbar].
\end{equation}
This is consistent with the analogous result for the polarization
moment (\ref{Jpol}) in the Dirac--Yang-Mills theory, where one finds
$[P_{{}_K}]=[e]$ (with $e$ as the non-Abelian charge or the usual
electric charge in the Abelian case). Dimensionwise, the
``translational charge'', which defines the translational moment, is
thus a {\it momentum} and the ``Lorentz charge'' an {\it angular
  momentum}.

In a remarkable way, the gravitational moments are closely related to
the spin of a Dirac particle. The relocalization superpotential $U$
and the translational moment ${\check M}_\alpha$ both can be expressed
in terms of the convective spin alone via the {\it identities}:
\begin{equation}
U \equiv -\,\eta_{\mu\nu}\wedge{}^\ast\tau^{(c)\mu\nu},\qquad
{\check M}_\alpha\equiv -\,\eta_{\alpha\mu\nu}\wedge{}^\ast\tau^{(c)\mu\nu}.
\end{equation}
Since ${\check M}_\alpha$ and $\tau^{(c)\mu\nu}$ have the same number
of independent components (namely 24), the last algebraic identity may be
inverted, giving the convective spin current in terms of the translational
moment. 

At first sight, it may be unclear that the Lorentz moment
(\ref{lorgrav}) is also related to spin. However, let us consider its
square invariant:
\begin{eqnarray}
M_{\alpha\beta}\wedge{}^\ast M^{\alpha\beta}&=&\left({\hbar^2\over 8mc}
\right)^2\left[-(\overline{\Psi}\Psi)^2 + (\overline{\Psi}\gamma_5\Psi)^2
\right]\eta_{\alpha\beta}\wedge\vartheta^\alpha\wedge\vartheta^\beta\nonumber\\
&=& 3\left({\hbar^2\over 4mc}\right)^2(\overline{\Psi}\gamma_\alpha
\gamma_5\Psi)(\overline{\Psi}\gamma^\alpha\gamma_5\Psi)\,\eta \nonumber\\
&=&{1\over 2}
\tau_{\alpha\beta}\wedge{}^\ast\tau^{\alpha\beta}.\label{eggegg}
\end{eqnarray}
Here, in the first line, we used the representation (\ref{lorgrav1}).
Then we rearranged the products, which are bilinear in the spinor
fields, by means of the Fierz identity.  As we recognize, in contrast
to the translational moment, the Lorentz moment is directly related to
the {\it complete} Dirac spin $\tau_{\alpha\beta}$.
Note that in applying the Gordon decomposition non-zero mass
was assumed. In the limit of massless Dirac particles, the expression
in (\ref{eggegg}) would vanish.

Some further insight can be obtained if we calculate the gravitational
moments for specific spinor field configurations. The plane waves
\begin{equation}
\Psi=\Psi_0(p) e^{-{i\over \hbar} p_\alpha x^\alpha},\label{plane}
\end{equation} 
as general solution of the free Dirac equation, are very important in
this context.  Substituting them into (\ref{Mtil}) and
(\ref{lorgrav1}), we find:
\begin{eqnarray}
{\check M}_\alpha &=& p_\alpha\,{\hbar\over 2mc}\,
\overline{\Psi}_0{}^\ast\widehat{\sigma}\Psi_0,\label{Maplane}\\
M_{\alpha\beta} &=& \hbar \,{\hbar\over 8mc}\,
\overline{\Psi}_0\Psi_0\,\eta_{\alpha\beta}.\label{Mabplane}
\end{eqnarray}
The (reduced) Compton wavelength $\hbar/mc$ in
(\ref{Maplane})-(\ref{Mabplane}) evidently provides the correct
dimension for these two-forms. In deriving (\ref{Maplane}), we used the
identity which holds for any solution of the Dirac equation (i.e. when
eqs.\ (\ref{10-4.11a})-(\ref{10-4.11b}) are satisfied):
\begin{equation}
i\left(\overline{\Psi}\,^\ast\widehat{\sigma}\wedge D\Psi - 
\overline{D\Psi}\wedge\,^\ast\widehat{\sigma}\Psi \right) = 
{}^\ast D(\overline{\Psi}\Psi).\label{MD}
\end{equation}
Since for the plane waves (\ref{plane}) the scalar $\overline{\Psi}\Psi=
\overline{\Psi}_0\Psi_0$ is constant (standard normalization is then 
$\overline{\Psi}_0\Psi_0=(mc/\hbar)^3\,\,$), we immediately find that the
generating three-form (\ref{Udir}) vanishes, $U=0$. 

In the non-relativistic approximation, we get for the translational moment
(\ref{Maplane}):
\begin{equation} 
{\check M}_\alpha={p_\alpha\hbar\over m}\,dt\wedge S + {\cal O}\left(
{v\over c}\right), \qquad S:=\psi_0^\dagger\hbox{$\scriptstyle{1\over 2}$} 
\hbox{\boldmath $\sigma$}\psi_0\,d\hbox{\boldmath $x$}.\label{Manonrel1}
\end{equation}
This result is a complete analog of (\ref{NonRel1}) for the
non-relativistic polarization moment in Dirac-Yang-Mills theory.

At the same time, no further clarification of the structure of the Lorentz 
moment (\ref{Mabplane}) occurs in the non-relativistic limit. In particular, 
it is {\it not} proportional to the spin one-form $S$, unlike the translational
moment ${\check M}_\alpha$ and the polarization current $P_{{}_K}$. 

\section{Discussion}

Inherent in the structure of the ``inertial currents''
$\bigl(\Sigma_\alpha,\tau_{\alpha\beta}\bigr)$ of the Dirac field,
namely of energy-momentum and spin, is the existence of convective and
polarization pieces, the latter ones being exterior covariant
derivatives of the gravitational moments $\bigl(\check
M_\alpha,M_{\alpha\beta}\bigr)$ of translational and Lorentz type,
respectively. This discovery is made on the level of
special-relativistic field theory, i.e., in (flat) Minkowski
spacetime. In other words, we were able to identify the gravitational
moments of the Dirac field {\it without} any involvement of the
gravitational field itself.  Rather, we only assumed that the
canonical {\it energy-momentum} and {\it spin} currents are the {\it
  sources} of gravity. With this assumption, which is in accord with
the Einstein-Cartan theory of gravity, we can tell from our results
that the field strengths of gravity have to be represented by two
two-forms with the following structure:
$\bigl(T_\alpha,R_{\alpha\beta}=-R_{\beta\alpha}\bigr)$.  In a future
paper we will show, as a final proof of our conception, that the
moments $\bigl(\check M_\alpha,M_{\alpha\beta}\bigr)$ couple, indeed,
to the field strengths $\bigl(T_\alpha,R_{\alpha\beta}\bigr)$,
provided the field strengths are interpreted as torsion and curvature
of spacetime.

\section{Acknowledgments}

This work was partially supported by CONACyT, grant No. 3544--E9311,
and by the joint German-Mexican project KFA-Conacyt E130-2924 and
DLR-Conacyt 6.B0A.6A. Moreover, EWM acknowledges support by the
short-term fellowship 9616160156 of the DAAD (Bonn) and YNO by the
project He 528/17-2 of the DFG (Bonn).

\section{Appendix}

Our general notation is as follows: The spacetime is four--dimensional
with a metric $g$ of signature $(+,-,-,-)$. A local frame is denoted
by $e_\alpha$ ($\alpha=0,1,2,3$; $\Xi=1,2,3$) and the dual coframe by
$\vartheta^\alpha$.  They fulfill the relation $e_\alpha \rfloor
\vartheta^\beta=\delta^ \beta_\alpha$, with $\rfloor$ denoting the
interior product.  For a holonomic or coordinate basis, we have
$d\vartheta^\alpha=0$; then there exists a local coordinate system $\{
x^i=(x^0,x^a) \}$ ($i=0,1,2,3$; $a=1,2,3$) such that $e_\alpha=
\!\!\!\!\!{}^\ast\, \delta_\alpha^i \partial_i$ and
$\vartheta^\alpha=\!\!\!\!\!{}^\ast\,\delta_i^\alpha dx^i$.
Anholonomic indices are always taken from the Greek and holonomic
indices from the Latin alphabet.  The Hodge star operator is denoted
by ${}^\ast$. Let $\eta:= {}^\ast 1$ be the volume four-form. The
following forms span the exterior algebra at each point of spacetime:
\begin{eqnarray}
\eta_\alpha&:=&e_\alpha\rfloor\eta={}^\ast\vartheta_\alpha,\\
\eta_{\alpha\beta}&:=&e_\beta \rfloor \eta_\alpha={}^\ast(\vartheta_\alpha
\wedge\vartheta_\beta),\\
\eta_{\alpha\beta\gamma}&:=&e_\gamma \rfloor \eta_{\alpha\beta}=
{}^\ast(\vartheta_\alpha\wedge\vartheta_\beta\wedge\vartheta_\gamma)\,,\\
\eta_{\alpha\beta\gamma\delta}&:=&e_\delta\rfloor\eta_{\alpha\beta\gamma}=
{}^\ast(\vartheta_\alpha\wedge\vartheta_\beta\wedge\vartheta_\gamma\wedge
\vartheta_\delta).
\end{eqnarray}

For the flat metric of Minkowski spacetime, $o_{\alpha\beta}= {\rm
  diag}(+1,-1,-1,-1)$, we choose the Dirac matrices in the form:
\begin{equation}
\gamma^{\hat{0}}=\left(\begin{array}{cr}\hbox{\boldmath $1$}&
\hbox{\boldmath $0$}\\\hbox{\boldmath $0$}&-\hbox{\boldmath $1$}
\end{array}\right),\quad
\gamma^a=\left(\begin{array}{cc}\hbox{\boldmath $0$}&\sigma^a\\ -\sigma^a 
&\hbox{\boldmath $0$}\end{array}\right),\quad
a=1,2,3.\label{gammas} 
\end{equation}
Here $\sigma^a$ are the standard $2\times 2$ Pauli matrices. Two important 
elements of the Dirac algebra are:
\begin{equation}
\widehat{\sigma}_{\alpha\beta}:= \frac{i}{2}
(\gamma_\alpha\gamma_\beta-\gamma_\beta\gamma_\alpha)\,,\label{sigD}
\end{equation}
\begin{equation}
  \gamma_5:= -{i\over
    4!}\,\eta_{\alpha\beta\mu\nu}\,\gamma^\alpha\gamma^\beta
  \gamma^\mu\gamma^\nu
  =-i\,\gamma^{\hat{0}}\gamma^{\hat{1}}\gamma^{\hat{2}}
  \gamma^{\hat{3}} =\left(\begin{array}{rr}\hbox{\boldmath $0$}&
      -\hbox{\boldmath $1$}\\-\hbox{\boldmath $1$}&\hbox{\boldmath
        $0$}
\end{array}\right).\label{gamma5}
\end{equation}
It is convenient to convert the constant $\gamma_\alpha$ matrices 
into Clifford algebra-valued one- or three-forms, respectively:
\begin{equation}
  \gamma:=\gamma_\alpha\,\vartheta^\alpha\,,\qquad{}^\ast\gamma=\gamma^\alpha
  \,\eta_\alpha\,.\label{10-4.2}
\end{equation}
Correspondingly, we obtain a two-form:
\begin{equation}
\widehat{\sigma}:=\frac{1}{2}\,\widehat{\sigma}_{\alpha\beta}\,
\vartheta^\alpha\wedge\vartheta^\beta=\frac{i}{2}\,
\gamma\wedge\gamma\,.\label{10-4.4}
\end{equation}
Two important identities hold for these Clifford algebra-valued
objects:
\begin{eqnarray}
\widehat{\sigma}_{\alpha\beta}\,{}^\ast\widehat{\sigma}&=& \eta_{\alpha\beta}
- i\,\gamma_5\,\vartheta_\alpha\wedge\vartheta_\beta 
- 2i\,\vartheta_{[\alpha}\wedge e_{\beta]}\rfloor{}^\ast\widehat{\sigma} 
,\label{sisi3}\\
{}^\ast\widehat{\sigma}\,\widehat{\sigma}_{\alpha\beta}&=& \eta_{\alpha\beta}
- i\,\gamma_5\,\vartheta_\alpha\wedge\vartheta_\beta 
+ 2i\,\vartheta_{[\alpha}\wedge e_{\beta]}\rfloor{}^\ast\widehat{\sigma} 
.\label{sisi4}
\end{eqnarray}

\end{document}